\documentclass[preprint]{elsarticle}
\usepackage[utf8]{inputenc}
\usepackage{xcolor}
\usepackage{amssymb}
\usepackage{url}

\definecolor{ourblue}{HTML}{4f81bd}

\begin{document}
\begin{frontmatter}

\title{Semantics between customers and providers:\\
The relation between product descriptions, reviews,\\
and customer satisfaction in E-commerce}

\author[label2]{Carlos A. Rodriguez-Diaz}
\author[label1]{Sergio Jimenez}
\author[label1]{Daniel Bejarano}
\author[label1]{Julio A. Bernal-Ch\'{a}vez}
\author[label2]{Alexander Gelbukh}
\address[label1]{Instituto Caro y Cuervo, Bogot\'{a}, Colombia.}
\address[label2]{Instituto Polit\'{e}cnico Nacional, Centro de Investigación en Computaci\'{o}n,\\Mexico City, Mexico.}
\begin{abstract}
In social commerce, users dialogue with each other on the topics related to the providers' products. However, the language customers use may vary from the language vendors use on their e-commerce websites and product descriptions. This situation can lead to possible misunderstandings in the social dialogue between customers, and incidental costs in the dialogue between customers and vendors. One possible reason for this difference is that words used by customers may have different meanings compared to those used by product description writers. We present a novel approach to measure this potential lexical-semantic gap for various e-commerce domains using an information-theoretical approach based on a large corpus of user reviews and product descriptions. Additionally, we use neural word embeddings to identify words with the highest semantic drift between reviews and descriptions as a tool to construct a ground truth for the task. We found that low levels of lexical-semantic gap are related to better customer satisfaction. Besides, this work is a step towards a better understanding of the effect of the use of language in e-commerce and social commerce. Potential applications of this technology could lead to better communication between customers and improvements in customer satisfaction, services, and revenue. Furthermore, this study opens up perspectives for applications in other domains with relationships beyond client-vendor, such as citizen-government and patient-healthcare system.
\end{abstract}


\end{frontmatter}

\section{Introduction}

E-commerce is \textcolor{black}{an Internet} business mediated by a dialogue \textcolor{black}{about products} between suppliers and customers. Initially, this dialogue is one-way, as the supplier provides the customer with multimedia information about the description of the products, which customers use to make their purchasing decisions and set their expectations  \cite{kim2002critical}. In turn, in social e-commerce, users dialogue with each other to share experiences related to the products and services they bought, and the topics of conversation are usually based on the information provided by the supplier \cite{hajli2019impact}. In these dialogues, written language plays a critical role in product descriptions, user reviews, user-to-user conversations, and customer service.

The role of language in e-commerce can be compared to \emph{Plain Language}, an area dedicated to reduce the linguistic gap between public administration and citizens \cite{adler2012plain}. Plain Language problems are very similar to those that arise between customers and suppliers. First, citizens are a much larger population than those who write public communications. Second, specialized language is commonly used since administrators, lawyers, or politicians are in charge of communicating with citizens. Third, the sociodemographic and language differences between the two groups greatly influence the \textcolor{black}{effectiveness} of the communication process. The same motivations that make clear language desirable in public administration may apply to the e-commerce scene.

It is generally assumed that customers fully understand the information provided by the supplier, but there are several reasons why this might not be true. First, the user population is generally much larger and more diverse than the population that writes product descriptions. Second, depending on each domain, the product descriptions contain specialized language to a different degree that users may not  fully understand \cite{vengadasamy2004characteristics}. Third, due to the trend towards globalization of trade, the sociodemographic and linguistic differences between customers and suppliers can be considerable \cite{bingi2006challenges}. These factors and others, such as the inherently ambiguous nature of the language, the intensive use of automatic translators \cite{barrera2016enhancing}, among others, suggest that this problem may  significantly affect e-commerce. However, this problem has received little attention from the scientific community.

There are mainly two situations where this possible gap between the language of the users and the suppliers generates problematic situations. First, if \textcolor{black}{the} user's expectations \textcolor{black}{about} a product are wrong due to this gap, customer satisfaction can be compromised. Second, users who interact in social commerce can reach misunderstandings that threaten the effectiveness and benefits of that social mechanism. Identifying and characterizing this gap is the first step to provide possible solutions to alleviate these problematic situations. 

From a marketing perspective, the relationship between customer satisfaction and business benefits has long been established \citep{rust1993customer} and recently reconfirmed in e-commerce scenarios \citep{eid2011determinants,tripathi2014customer}. Theories aimed at modeling and explaining customer satisfaction are mainly based on economic and behavioral factors related to users and business \citep{rese2003relationship}. On the other hand, empirical studies have related customer satisfaction with factors such as service quality \citep{kassim2010effect,biswas2019influence}, logistics \citep{hua2015empirical,skordoulis2018commerce}, and client's expectation fulfillment \citep{gong2010research}. The linguistic aspects of customer satisfaction in e-commerce have been studied from the perspective of analyzing the content of user reviews, that is, what words and expressions represent satisfaction or dissatisfaction in customer discourse \citep{mu2021research}. This user-centric approach is most useful for customer profiling and product development. However, few studies have studied the language usage from the side of providers from a holistic perspective and the possible factors related \textcolor{black}{with customer satisfaction}, such as semantics, style, and complexity \citep{pryzant2017predicting,elsholz2019exploring}. Given that the use of language by customers is only an observable factor, but the use of language on the provider's side is a relatively controllable factor, the possible relationships of the latter with client satisfaction would open managerial perspectives to improve business profits.

From a linguistic point of view, \emph{semantic drift} is the change or difference in the meaning of words between two populations differentiated by usage, dialect, time, or demography \citep{beinborn2020semantic}. A word with semantic drift in social commerce has a predominant sense of use by customers that is different from its meaning in the textual content produced by the supplier. For example, in the domain of sports products, for customers, the word ``core'' refers to a set of muscles of the human body around the torso, while in product descriptions, the predominant meaning refers to internal the material or component from which a particular product is made. While both parties may be aware of the two meanings, and both may disambiguate depending on the contexts \cite{piantadosi2012communicative}, if the preferred meanings differ significantly, it is plausible that this word is a factor for misunderstandings. Also, if many words in a domain exhibit this behavior, the quality of written communication can be compromised.

Manual identification of words with semantic drift in a particular domain is an overwhelming task for linguists and lexicographers due to the variety and specificity of the domains and the large number of texts that must be analyzed. Concordance analysis allows linguists to compare the contexts of use of a word in two sets of texts. However, even using concordancing software, since a community's vocabulary size  is often large, detecting words with semantic drifts is a tedious, overwhelming, and expensive task.

Alternatively, automatic approaches have been studied to identify and measure semantic drift. These are based on the idea of determining the meaning of the words by the linguistic distributional hypothesis that states, ``you shall know a word by the company it keeps'' (\citet{firth1957synopsis}). For this, it is usually required that each of the speech communities is represented by a corpus of considerable size to be able to identify statistically significant differences between the contexts of use of a word \cite{roberts2016assessing}. The traditional methods for this are the vector space model \cite{salton1975vector}, random indexing \cite{sahlgren2006word}, latent semantic indexing \cite{deerwester1990indexing}, and others, which have proven to be relatively effective but have limitations in vocabulary length and in the number of texts that they support. Recently, there was a paradigm shift in natural language processing towards  neural networks for semantic analysis of words \cite{turian2010word}. In particular, \emph{words embeddings} \cite{mikolov2013distributed,pennington2014glove} make it possible to obtain vector representations for words in geometric spaces that have semantic properties of similarity, analogical and compositional reasoning. Word embeddings overcome the limitations of the previous methods, allowing large amounts of text (billions of words) to be processed efficiently.

The textual data used in this study seeks to be representative of the e-commerce and social commerce scenarios. Customer-written product reviews are the basic mechanism by which the customer community shares information about their shopping experiences \cite{zheng2013capturing}. We will use these reviews assuming that other types of communication between customers in social commerce have the same linguistic features.  We use the product's textual descriptions to represent the suppliers' language, assuming that they share the same linguistic features \textcolor{black}{with} other textual components in the user interfaces and other communications between customers and vendors \cite{pryzant2017predicting}.

However, there is a significant limitation in the social commerce scenario for applying approaches based on word embeddings. While the amount of textual content users produce in their social interaction is large and suitable for this approach, the amount of text that describes the products is much smaller. This imbalance makes the vector representations of the words in the customer corpus of good quality\textcolor{black}{,} while those obtained from the product descriptions \textcolor{black}{produce} noisy \textcolor{black}{vectors}.

To address this challenge, in this study, we present a new approach based on information theory \cite{shannon1948mathematical} to measure the semantic gap between two corpora highly unbalanced in size. We hypothesize that by using a commonly used file compression algorithm (i.e., bzip2) and a randomization mechanism, it is possible to combine user texts and descriptions so that their compressed sizes can be used to measure the semantic gap. To assess the validity of this method, we apply word embeddings to the large corpus representing customers (product reviews) and the small corpus of product descriptions for a particular domain. With this, we obtain a set of words with considerable semantic drift but with a very high error rate due to the marked imbalance in size. We then manually identify the few actual words with semantic drift in that set by analyzing the closest neighbor words. These manually selected words and the magnitude of their semantic drift constitute a ground truth for evaluating the proposed compression-based method. 

Additionally, this new method was tested on a large product description and review dataset covering 28 product domains. In these experiments, we explore the possible relationship between the semantic gap and customer satisfaction.

The rest of the article is organized as follows: \textcolor{black}{In Section \ref{sec:litrev}, we present a review of some related work to the analysis of product reviews and descriptions, and their relationship to customer satisfaction.} Besides, we provide some background concepts aimed at readers unfamiliar with topics of computational linguistics to understand the proposed methods and hypothesis better. In Section \ref{sec:method} we present these methods and the data that were used. In Section \ref{sec:results}, we present the results obtained, and the evaluations carried out. Then in Section \ref{sec:discussion}, we present the discussion of the results. Finally, in Section \ref{sec:conclusions}, we present the conclusions derived from this study.

\section{Literature Review}\label{sec:litrev}

This section seeks to contextualize this study as follows. First, in subsection \ref{sec:related_work} we review those previous works that have studied the possible relationships between the textual content of product descriptions and user reviews, with commercial factors such as customer satisfaction and profit. Then, in subsections \ref{sec:corpus_compression} and \ref{sec:drift_embeddings} we provide contextualization for the methods presented in Section \ref{sec:method} (Materials and Methods) thus allowing a better understanding of those methods and the hypotheses of this study. Finally, in subsection \ref{sec:hypothesis} such hypothesis are presented.

\subsection{Related Work}\label{sec:related_work}
\citet{francis2002pirqual} verified in a survey of electronic commerce customers that the quality of the information in the description of the attributes of the products positively influences customer satisfaction. In a similar study, \citet{ludin2014factors} found a significant positive relationship between the quality of product information and customer satisfaction, which implies a transitive relationship with e-loyalty. They concluded that the quality of the information helps customers make better purchasing decisions that, incidentally, lead to increased customer satisfaction. The recent study by \citet{patrada2021effect} reconfirms these results, showing the undeniable effect that the quality of product information has on customer satisfaction. We assume that among the product information there is always textual content where quality means a well-written, informative, concise and coherent text. \citet{mou2019impact} also found that the quality of written descriptions also has positive relationships with cognitive involvement, affective product involvement, and platform customer involvement. In conclusion, the quality of written content related to products has a positive effect on e-commerce taking as evidence the opinions of customers.

\textcolor{black}{In the field of artificial intelligence, \citet{pryzant2017predicting} provided evidence of the close relationship between product descriptions and sales. \textcolor{black}{To do this, they trained a deep learning neural network from the sequences of words in the product descriptions to identify the lexical patterns associated with sales.} They found that even controlling for confounding factors like product type, brand, and price, textual descriptions are highly predictive of
sales ($R^{2}$ around 0.80). Although this is not proof of a causal relationship, this corroborates previous results in the field of marketing, where the positive effect of the quality of product descriptions is already known. Furthermore, evidence is added that not only the overall quality but also the lexical and probably semantic content of these descriptions is responsible for the possible effects on sales and customer satisfaction.} 

Other studies have addressed the interaction between product descriptions and customer reviews. \citet{wang2018extracting} found that for some product characteristics in the descriptions, it is possible to automatically find related emotional factors in the reviews to develop products with affective responses. Motivated by the problem of shortage or poor quality of product descriptions in some domains, some studies have tackled the task of creating product descriptions from customer reviews automatically \citep{novgorodov2019generating,zhang2019automatic,chen2019towards}. In addition to solving the problem posed, the product descriptions turn out to be \textcolor{black}{a} reliable content for customers derived from crowd-sourced material. Other studies have even proposed the automatic generation of personalized product descriptions in order to optimize the effect of their textual content on customer satisfaction \citep{elad2019generating}. However, although the motivations for these studies are plausible, these initiatives do not provide empirical evidence that these synthetic product reviews have an effect on customer satisfaction.

In this context, this study seeks to improve the understanding of the semantic textual features in product descriptions and their possible effects on customer satisfaction. In particular, we will study the degree of semantic alignment between the meanings of words used in product descriptions and in reviews. This is a meta-linguistic characteristic that is independent of the used language or its vocabulary. Besides, this feature is associated with a universal of human language that is ambiguity. In contrast, previous work has identified particular words in the English language, in particular domains, that are predictive features of sales or satisfaction. These results are difficult to generalize to other languages or domains. Our contribution is to provide a general method to measure the semantic gap between descriptions and reviews, and thereby empirically test the hypothesis that it is related to customer satisfaction.


\subsection{Corpus Information by Compression}\label{sec:corpus_compression}

\citet{cilibrasi2003clustering} showed that the Kolmogorov complexity of a data file, which is the size of its theoretical minimum compressed version, can be obtained with a normal compressor. Said compressor $C$, should fulfill the properties of \emph{idempotency} $C(xx)=C(x)$, \emph{monotonicity} $C(xy)\geq C(x)$, \emph{symmetry} $C(xy)=C(yx)$, y \emph{distributivity} $C(xy)+C(z)\leq C(xz)+C(yz)$. Most file compressors (e.g., zip, rar, bzip2, among others) asymptotically approximate the properties of a normal compressor and hence Kolmogorov complexity. That way, this notion of information is associated with the size of the compressed data file, applying in principle to any type of data, but it is handy for large textual corpora.

A file compressor takes a sequence of numerical data and looks for patterns to recognize redundancy and somehow ``understand'' the sequence so that it can be reproduced again from a condensed version. Thus, a sequence of random numbers is not compressible, while any sequence with some notion of order is compressible to some extent. The idea of measuring the information in a corpus by its compressed size is useful for measuring the amount of information in some other feature of the corpus. For example, suppose that $x$ is a textual corpus of English text with $|x|$ bytes in size and $C(x)$ bytes in its compressed version. Now, let's create a variant $x'$ by replacing all occurrences of the word ``that'' with ``this''. The uncompressed size $|x|$ is equal to $|x'|$, but $C(x)>C(x')$ because $x$ has more information than $x'$ because it has one less word in its vocabulary. The difference $\delta=C(x)-C(x')$ represents the amount of information that conveys the word ``that'' in corpus $x$, which is the feature to be measured in this example.

\citet{montemurro2011universal} used the same compression principle to measure the amount of information associated with the order of words in a corpus. They produced a variant of an original corpus by shuffling the words in each sentence, making two corpora of identical size. The difference between the compressed sizes of these two corpora represents the amount of information associated with the order of the words. They found that this difference is universal in different language families. Similarly, the method presented in this study manipulates the text corpora produced by the speaking communities of customers and providers in such a way that the difference in size between the two versions of the corpora represents the semantic gap between the two communities.

\subsection{Measuring Word Drifts by Word Embeddings}\label{sec:drift_embeddings}
Neural word embeddings are a set of methods for obtaining vector representations for words in a high-dimensional Euclidean space. The input is a large textual corpus; the output is a  $n\times m$ matrix, where each row is a vector of dimension $m$ associated with each of the \textcolor{black}{$n$} words in the vocabulary of the corpus. Thus, each word is located in some coordinates in the space of $m$-dimensions, where its position has semantic properties. The main property of that space is that the distance between the words represents a value of similarity or semantic relationship. Therefore, two words like ``car'' and ``automobile'' should be close, while ``car'' and ``cucumber'' should be far from each other. Compositionality is another property of said semantic space, so the vector sum of the representations of, say, ``doctor'' and ``heart'' must be close to ``cardiologist''. Also, this space allows for analog reasoning. For example, the relative distance and direction between the words ``king'' and ``queen'' are the same as those between ``prince'' and ``princess''. This means that it is possible to solve analogies like ``king''$\rightarrow$``queen'': ``prince''$\rightarrow X$. These word representations are obtained from the weights of the links between neurons in an artificial neural network trained in a word co-occurrence prediction task using a large corpus as a data set for training.

 The semantic drifts of words can be determined by geometrically comparing the vectorial representation of the words obtained using two large corpora representing each community of speech or dialect. The detection of word drifts, as well as the complementary task of finding semantic equivalences, can be performed using word embeddings \citep{ruder2019survey}. For these tasks, the input is two corpora, each representing a community of speech or dialect, and the output is a single semantic Euclidean space where the words to be tested are labeled with an identifier of their source corpus. The distances between pairs of labeled words in such space are compared.  Words that are close to its pair are considered semantic equivalences, and the farthest pairs are considered semantic drifts. Approaches for obtaining such single semantic space from two corpora range from geometric alignment \citep{artetxe2018robust}, conditional probability \cite{han2018conditional,kulkarni2015statistically}, and meta-algorithms on current word embedding algorithms \cite{gouws2015simple,vulic2015bilingual}. 
 
To evaluate the quality of the results of the words identified with semantic drift, it is necessary to have a gold standard, which generally must be constructed manually. Professional linguists typically perform this task and usually produces few instances \cite{frermann2016bayesian}. For example, in the DIACR-Ita challenge \citep{basile2020diacr}, there were only 18 words available in the gold standard to detect semantic drift in a diachronic corpus (that is, a corpus from the past and one from the present). In this challenge, \citet{angel2020nlp} showed that it is not necessary to obtain a single semantic space to identify semantic drifts but that this can be done using the word embeddings obtained separately for each corpus. To identify if a word has semantic drift, its corresponding sets of $k$ neighboring words in each semantic space are collected and compared with the Jaccard index. The target words that get the lowest value of this index have the highest semantic drift. This approach obtained good results, and it was robust to the imbalance between corpus' sizes with few texts from the past and many texts from the present. In the method presented in this study, we will use that idea to alleviate the problem of imbalance between the amounts of text produced by customers compared to product descriptions.

\subsection{Research Questions and Hypothesis}\label{sec:hypothesis}
This study aims to propose an adequate way to measure the semantic gap between the language of customers and suppliers in an e-commerce scenario, and to identify whether such a gap exists and what is its magnitude for different market domains. In addition, it seeks to determine whether this situation is related to customer satisfaction. 

The hypotheses that we seek to test in an e-commerce environment, where user reviews and product descriptions are available, are the following:\\
\begin{itemize}
    \item$H_{1}$: If there are differences in the word usage contexts (i.e., senses) between reviews and descriptions, then the difference in the amount of information between them due to that (information theoretically speaking) is a measure of the magnitude of the semantic gap between customers and suppliers.\\
    
    \item To operationalize the $H_1$ hypothesis in the proposed theoretical framework, we propose the following null auxiliary hypothesis. Assume that you have two corpora: the first (called $True$) is a combination of revisions and descriptions, and the second (called $Rand$) is a copy of $True$ to which random changes are introduced in the contexts of use of words.\\
    
    $H_0^{aux}$: If there is no size difference between the compressed files of $True$ and $Rand$, then there are no differences between the senses of use of the words in reviews and descriptions.\\

    \item$H_{2}$: The greater the lexical-semantic gap between customers and providers, the lower the customer satisfaction.
\end{itemize}

\section{Materials and Methods}\label{sec:method}

The methodology presented in this section aims to address two objectives. First, present a method to measure the degree of semantic gap between the language of customers and suppliers. Second, provide the framework for checking the hypothesis whether the magnitude of this gap is related to customer satisfaction in social e-commerce environments. These objectives require a considerable amount of representative texts, where a domain level categorization is available, to be able to observe and analyze the plausibility of possible variations. In addition,  to validate the hypothesis, it is necessary to have associated metadata that provide quantitative information on customer satisfaction (for example, ratings given by customers). The main methodological challenge is the design of a semantic gap measure in highly unbalanced data such as the huge textual output of customers in the form of product reviews, in contrast to the few quantities of product descriptions available. Since the known methods for this (i.e. neural word embeddings) require a fairly balanced corpus, we propose a measure of the semantic gap based on information theory. To validate this method, we applied neural word embeddings, which are widely recognized for their suitability for the task, on the unbalanced data and performed manual curation of the results to build a gold standard for the task. Thus, when comparing this gold standard with the results of the proposed method, a completely automatic method for the task would be validated and obtained. Finally, the gap measures will be analyzed qualitatively and then quantitatively compared to customer satisfaction data. 

\subsection{The Data}
We used the \emph{Amazon Review Data 2018} \cite{ni2019justifying} that contains around 180 million of reviews and 12 million product descriptions distributed along 28 domains in power-law like. We used the text of the reviews and extracted text from the product data fields \emph{title}, \emph{tech1},  \emph{description}, \emph{feature}, and \emph{similar item}. Each of the reviews contains a rating on a discrete scale between 5 (best) to 1 (worst) that we will use as a quantitative indicator of customer satisfaction. Sentence boundaries were identified using the NLTK\footnote{Natural Language Toolkit at \url{https://www.nltk.org/}}, and rows from tabular descriptions were considered as ``sentences''. Next, all non-alphabetical tokens were removed. The sizes of the resulting review and description corpora per domain are shown in Table \ref{corpus_data}.

\begin{table}
\caption{\label{corpus_data} Number of words for each domain and word count ratio between reviews and descriptions}

{\small
\begin{tabular}{p{4cm}rrp{2.0cm}}
\hline
\textbf{Category}  & \textbf{Reviews' words} & \textbf{Descriptions' words} & \textbf{Word imbalance ratio} \\
\hline
All Beauty                & 13,454,743          & 1,285,534                  & 10:1                        \\
Amazon Fashion            & 22,235,491          & 5,919,402                  & 4:1                         \\
Appliances                & 19,211,478          & 4,566,216                  & 4:1                         \\
Arts Crafts \& Sewing      & 74,693,032          & 27,578,534                 & 3:1                         \\
Automotive                & 226,834,970         & 161,065,374                & 1:1                         \\
CDs \& Vinyl               & 386,158,660         & 21,973,448                 & 18:1                        \\
Cell Phones \& Accessories & 349,903,420         & 72,093,498                 & 5:1                         \\
Clothing Shoes \& Jewelry  & 866,541,485         & 329,871,943                & 3:1                         \\
Digital Music             & 55,810,053          & 1,554,864                  & 36:1                        \\
Electronics               & 1,024,197,307       & 135,120,799                & 8:1                         \\
Gift Cards                & 2,372,944           & 84,021                     & 28:1                        \\
Grocery \& Gourmet Food    & 149,018,135         & 26,804,856                 & 6:1                         \\
Home \& Kitchen            & 759,164,935         & 192,811,582                & 4:1                         \\
Industrial \& Scientific   & 54,799,737          & 16,627,174                 & 3:1                         \\
Kindle Store              & 382,079,930         & 3,420,666                  & 112:1                       \\
Luxury Beauty             & 21,696,560          & 1,040,486                  & 21:1                        \\
Magazine Subscriptions    & 3,610,776           & 133,637                    & 27:1                        \\
Movies \& TV               & 469,363,734         & 18,061,999                 & 26:1                        \\
Musical Instruments       & 69,072,246          & 18,739,313                 & 4:1                         \\
Office Products           & 191,650,532         & 49,943,042                 & 4:1                         \\
Patio Lawn \& Garden       & 186,284,611         & 45,539,118                 & 4:1                         \\
Pet Supplies              & 269,510,943         & 32,330,942                 & 8:1                         \\
Prime Pantry              & 10,415,289          & 1,048,980                  & 10:1                        \\
Software                  & 33,571,934          & 4,916,812                  & 7:1                         \\
Sports \& Outdoors         & 468,858,744         & 91,971,650                 & 5:1                         \\
Tools \& Home Improvement      & 327,711,025         & 107,013,154                & 3:1                         \\
Toys \& Games              & 273,351,609         & 82,249,226                 & 3:1                         \\
Video Games               & 179,005,493         & 11,711,784                 & 15:1     \\
\hline
\end{tabular}
}
\end{table}

\subsection{Semantic Gap by Compression}\label{sec:semantic_gap}
Inspired by \citet{montemurro2011universal}, for each domain, we produced two versions of the same size of the combined reviews and description corpora, one with the original sentences (\emph{True}) and the other with random word shifts (\emph{Rand}). These shifts are designed so that the difference between the compressed sizes of the two corpora may reflect the semantic gap between customers and product description writers.  In computational linguistics, two words can be considered synonyms if they are interchangeable in many contexts \cite{miller1995wordnet}. The \emph{Rand} corpus represents the null hypothesis that a particular word $w$ (or a set of words) has the same (or close) meaning in reviews and descriptions because random exchanges should not affect the null hypothesis of interchangeability. If there is little or no difference in the compressed sizes, the null hypothesis is tested.

First, for each domain, we create a mixed corpus of reviews and descriptions by inserting a sentence from each source one at a time. The motivation for this is that most current compressors are stream-based, so they would incrementally encode the reviews and descriptions data, favoring the distributivity property of compression (see subsection \ref{sec:corpus_compression}). Since the sentences in the corpus of reviews are more abundant than in the corpus of descriptions, we start over with the first description as many times as necessary once we run out of descriptions. Next, we select the vocabulary $W$ to be tested for drift, removing from the common vocabulary between reviews and descriptions the 500 most frequent words and the words that appear less than 50 times. The motivation for this is that the most frequent words are usually functional words that have little or no ambiguity. Similarly, at the other end of the frequency spectrum, words with few occurrences provide little evidence for automatic analysis.

\begin{figure}	
\centering
\includegraphics[trim=0 280 218 0, clip,width=8 cm]{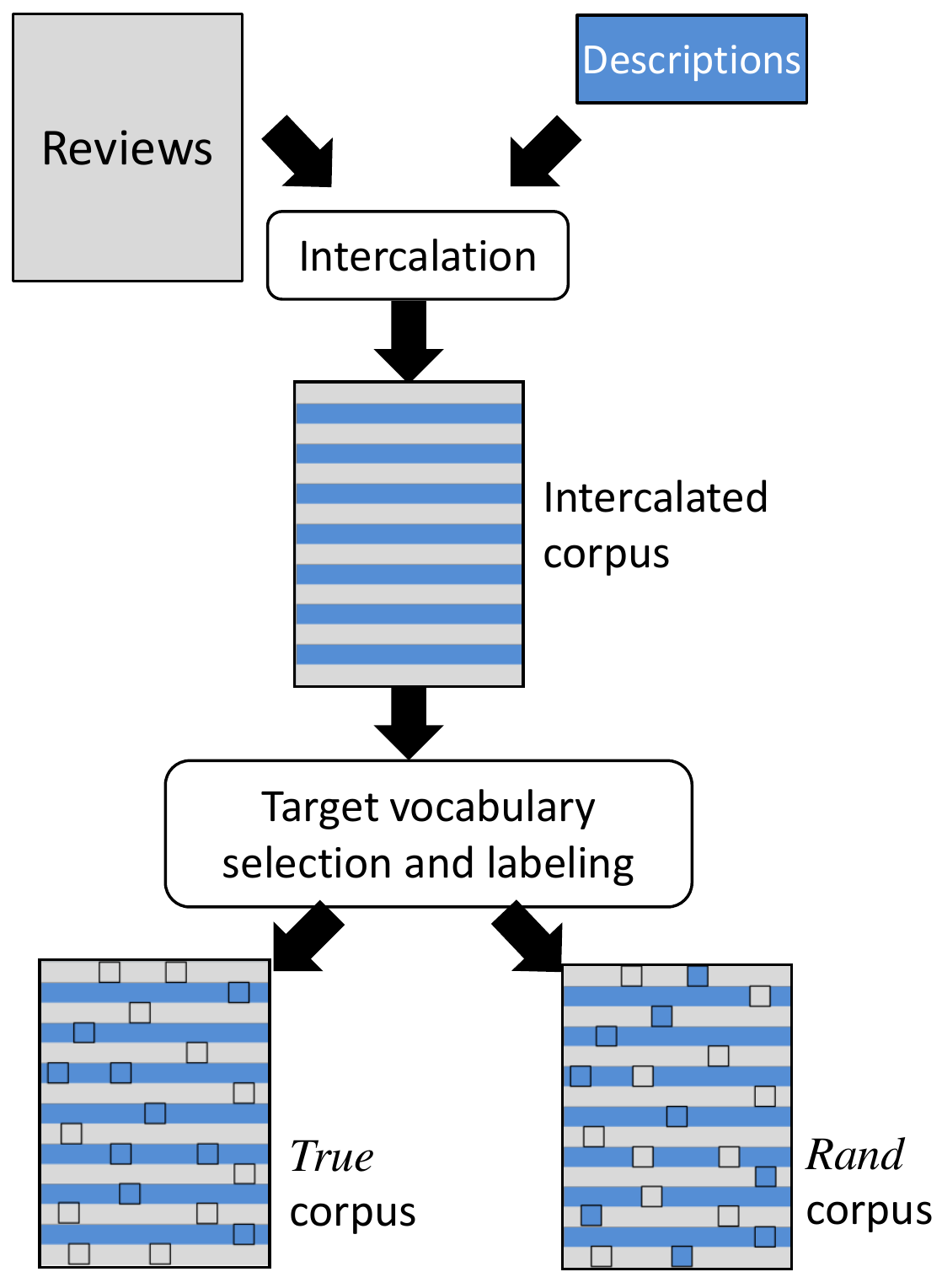}
\caption{The process to build the \emph{True} and \emph{Rand} corpus (of equal size) from the product reviews and descriptions \label{fig:corpus_manipulation}}
\end{figure}

Then for each word $w$ in $W$, we label each occurrence $w_R$ if $w$ occurs in a review and $w_D$ for a description. We call this the corpus \emph{True}. To produce the corpus \emph{Rand}, we randomly swap the labels $R$ and $D$ on each target word with a probability $P=0.5$ (i.e., by ``flipping a coin''). The justification for this value came from the analysis of the extreme values $P=0$, which means that there are no changes, and $P=1$, which means a tag exchange that does not produce changes either. The motivation for this randomization is that if $w$ has the same meaning in reviews and descriptions, then $w_R$ and $w_D$ can be swapped (i.e., interchanged in any context). On the contrary, if $w_R$ and $w_D$ have different meanings, the random exchange would place words in unexpected contexts that would increase the corpus's general information. Figure \ref{fig:corpus_manipulation} illustrates this process in a graphic analogy: gray represents text for reviews, blue for descriptions, horizontal stripes represent sentences, small squares are target words, and the fill color of these squares are their labels. Finally, we measure the size of the two compressed corpora using the \emph{bzip2}\footnote{\url{http://www.bzip.org/}} compressor, the same option that Wikipedia administrators use to store and distribute the backups of their encyclopedia.

Table \ref{corpus_example} shows an example of the \emph{True} and \emph{Rand} corpora for $w$ =``nature'' in the \emph{Electronics} domain. These corpora are the result of the successive alternation of a sentence of revisions and a sentence of descriptions. In the \emph{True} corpus, the tags of the target word `` nature '' match the source corpus of the sentence, while in the \emph{Rand} corpus the tags are randomly assigned. If the target word is used in the same sense in both corpora, most contexts would be similar to each other, and there would not be a considerable difference between \emph{True} and \emph{Rand}. Otherwise, the contexts would differ, and the random shifts would result in the tagged target word being placed in many inappropriate contexts, thus increasing the amount of information in the \emph{Rand} corpus. In the example in Table \ref{corpus_example}, only a single target word is considered, but this tagging process is done simultaneously for all target words, those that are neither very frequent nor very rare in the corpus. In particular, the word ``nature'' exhibits a drift in the \emph{Electronics} domain. The dominant meaning in descriptions is ``the outer world of living beings'', while in reviews, it is ``the basic character or quality''. However, in other domains such as \emph{CD \& Vinyl}, ``nature'' is related both in reviews and descriptions to idealism, spirituality, and serenity.

\begin{table}[ht]
\caption{Example of the \emph{True} and \emph{Rand} mixed reviews/descriptions corpora for the sample word ``nature''. A \colorbox{lightgray}{R} label is attached to target words from the corpus of reviews and  \colorbox{ourblue}{\textcolor{white}{D}} from descriptions.\\}
\renewcommand{\arraystretch}{1.5}
\small{
\begin{tabular}{p{0.9cm}|p{6.6cm}|p{0.9cm}|p{6.6cm}}
\hline
\textbf{True Label} & \textbf{True Corpus} & \textbf{Rand. Label} &\textbf{Random Corpus}  \\
\hline

 \colorbox{lightgray}{R}  & That is the nature\_\colorbox{lightgray}{R} of digital. & \colorbox{ourblue}{\textcolor{white}{D}} &  That is the nature\_\colorbox{ourblue}{\textcolor{white}{D}} of digital.\\
 
  \colorbox{ourblue}{\textcolor{white}{D}} & Improve your mood in nature\_\colorbox{ourblue}{\textcolor{white}{D}} euphony. & \colorbox{ourblue}{\textcolor{white}{D}} & Improve your mood in nature\_\colorbox{ourblue}{\textcolor{white}{D}} euphony.\\
  
 \colorbox{lightgray}{R} & The nature\_\colorbox{lightgray}{R} of wireless is noise. & \colorbox{lightgray}{R} & The nature\_\colorbox{lightgray}{R} of wireless is noise.\\
 
 \colorbox{ourblue}{\textcolor{white}{D}} & Bushnell nature\_\colorbox{ourblue}{\textcolor{white}{D}} view plus x compact binocular. & \colorbox{lightgray}{R} & Bushnell nature\_\colorbox{lightgray}{R} view plus x compact binocular.\\
 
 \colorbox{lightgray}{R} & Just the nature\_\colorbox{lightgray}{R} of the beast. & \colorbox{ourblue}{\textcolor{white}{D}} & Just the nature\_\colorbox{ourblue}{\textcolor{white}{D}} of the beast.\\
 
 \colorbox{ourblue}{\textcolor{white}{D}} & Timex cd clock radio with  nature\_\colorbox{ourblue}{\textcolor{white}{D}} sounds & \colorbox{lightgray}{R} & Timex cd clock radio with  nature\_\colorbox{lightgray}{R} sounds.\\
 
 \colorbox{lightgray}{R} & Love the retractable nature\_\colorbox{lightgray}{R} of this cord.& \colorbox{lightgray}{R} & Love the retractable nature\_\colorbox{lightgray}{R} of this cord.\\
 
 \colorbox{ourblue}{\textcolor{white}{D}} & Separate  nature\_\colorbox{ourblue}{\textcolor{white}{D}} pictures for each individual verse. & \colorbox{lightgray}{R} & Separate nature\_\colorbox{lightgray}{R} pictures for each individual verse.\\
 
 \colorbox{lightgray}{R}  & I rarely use the nature\_\colorbox{lightgray}{R} sounds. & \colorbox{ourblue}{\textcolor{white}{D}} & I rarely use the nature\_\colorbox{ourblue}{\textcolor{white}{D}} sounds.\\
 
 \colorbox{ourblue}{\textcolor{white}{D}} & Used and praised by   nature\_\colorbox{ourblue}{\textcolor{white}{D}} photographers. & \colorbox{lightgray}{R} & Used and praised by  nature\_\colorbox{lightgray}{R} photographers.\\
 
 \colorbox{lightgray}{R}  & That is just the nature\_\colorbox{lightgray}{R} of privacy filters. & \colorbox{ourblue}{\textcolor{white}{D}} &  That is just the nature\_\colorbox{ourblue}{\textcolor{white}{D}} of privacy filters.\\
 
\colorbox{ourblue}{\textcolor{white}{D}} & Perfect for nature\_\colorbox{ourblue}{\textcolor{white}{D}} sports and surveillance photos. & \colorbox{ourblue}{\textcolor{white}{D}} & Perfect for nature\_\colorbox{ourblue}{\textcolor{white}{D}} sports and surveillance photos.\\ 
 
 \hline
\end{tabular}
}
\label{corpus_example}
\end{table}


\subsection{Word Embeddings for Word Drift}\label{sec:sem_drift}

To detect words with different meanings between reviews and description corpora for each domain, we followed \citet{angel2020nlp} by training word embeddings for each corpus and comparing neighboring words for each target word. We use the implementation of \emph{word2vec} \citep{mikolov2013distributed} in Gensim \citep{rehurek_lrec} using the CBOW algorithm and a 5-word window (these are the default parameters). Since for all domains, each corpus of reviews is larger than its corresponding corpus of descriptions; we selected a vector size of 200 for reviews and 50 for descriptions to reduce the total number of parameters to learn for the embeddings from descriptions. Additionally, we only consider words that occur at least 50 times in the corpus of reviews and ten times in descriptions. Finally, two sets of word embeddings were obtained for each domain by training for five epochs (i.e., passes of the corpus) of the reviews corpus and ten epochs of the corpus of descriptions \textcolor{black}{(larger vector sizes usually should require more epochs)}.

Next, for each word in common between the two word embeddings for a domain, we obtained the $k=30$ closest neighbors in reviews and descriptions. All words in common were scored with the following expression:

\begin{equation}
S(w)=\log(\min(f_{w_r},f_{w_d}))\times (1-Jaccard(w_r,w_d))^p\label{eq:score}
\end{equation}

Here, $f_{w_r}$ is the frequency of the word $w$ in the reviews corpus and $f_{w_d}$ in the descriptions, $w_r$ is the set of the closest neighboring words of $w$ in the reviews' word embeddings and $w_d$ in descriptions, $J()$ is the Jaccard coefficient, and $p$ is a parameter for the multiplicative combination set to 5. \textcolor{black}{In computational linguistics it is used to logarithmically scale the frequencies of the words in a corpus due to Zipf's Law.} This scoring function aims to select the most frequent words in both corpora with few neighboring words in common.  It should be noted that the frequencies $f_{w_r}$ and $f_{w_d}$ are obtained from the artificially balanced corpus (\emph{True}) and not from the original corpus. That is, their values are comparable.

\subsection{Building a ground truth for the word drift task}\label{sec:groundtruth}

To provide the initial ground truth for finding the words with the highest semantic drift between reviews and descriptions, we rank the words in each domain according to $S()$ (eq. \ref{eq:score}) in descending order. Next, we identify the first ten words with semantic drift by manually analyzing the two neighboring word lists. The domains \emph{Gift Cards} and \emph{Magazine Subscriptions} were ignored due to the poor quality of their word embeddings from descriptions due to their small corpus size. However, we noticed that the semantic relationship of the $k$-nearest neighboring words is better in the word embeddings obtained from product reviews than that of descriptions. Many unrelated words had to be discarded from the set of neighbor words from descriptions to discern the meaning used in that corpus. A professional linguist performed this task. For most of the domains, the first ten words with a clear word drift were found among the first 200 words ordered by $S()$, thus obtaining 280 instances in the ground truth. Table \ref{tab:ground_truth} shows examples of these words for some domains. The entire manually created ground truth is available at \url{https://tinyurl.com/mhk5a4d4}.

\begin{table}[ht]
\caption{Examples of words exhibiting semantic drifts obtained from word embeddings trained separately on reviews and descriptions }\label{tab:ground_truth}
\renewcommand{\arraystretch}{1.5}
\begin{tabular}{p{20mm}lp{59mm}p{59mm}}
\hline 
{\small{}\textbf{Domain}}&{\small{}\textbf{Word}}&{\small{}\textbf{Neighbors in Reviews}} & {\small{}\textbf{Neighbors in Descriptions}}\\ \hline

{\small{}All Beauty} & {\small{}sexy} & {\small{}feminine classy elegant girly manly sparkly} & {\small{}lingerie tights underwear sleepwear lolita}\\

{\small{}Arts Crafts} & {\small{}friendly} &{\small{}efficient intuitive satisfying responsive} & {\small{}safe conscious biodegradable recyclable}\\

{\small{}Automotive} &{\small{}cool} &{\small{}badass sexy sweet cute nice snazzy neat classy} & {\small{}warm toasty refreshing cockpit blazing cozy}\\

{\small{}Arts Crafts  \& Sewing}&{\small{}hands}&{\small{} fingers finger wrist thumb fingertip} & {\small{}professionalism expertly handwash machine tlc}\\

{\small{}CDs \& Vinyl}&{\small{}lovers}&{\small{}enthusiast connoisseur fanatic fan collector} & {\small{}woman man lady fool baby letter girl kisses}\\

{\small{}Cell Phones \& Accessories}&{\small{}hard}&{\small{} difficult harder tough easy tricky} & {\small{}tpu silicone silicon hardshell tortoise}\\

{\small{}Clothing Shoes \& Jewelry
}&{\small{}left
}&{\small{} leaving cracked rubbed leaves corner} & {\small{} click  sizing  exploded  front length  forearm}\\

{\small{}Digital Music
}&{\small{}number
}&{\small{} handful slew couple several consecutive} & {\small{} top labels period shelf seam}\\

{\small{}Electronics
}&{\small{}bus
}&{\small{} buses train plane subway trains} & {\small{} nics uart quatech epp fsb}\\

{\small{}Kindle Store
}&{\small{}love
}&{\small{} adore loved enjoy loving loves} & {\small{} passion hope true trusts }\\

\hline
\end{tabular}
\end{table}

\begin{figure}[ht!]

\includegraphics[width=16 cm]{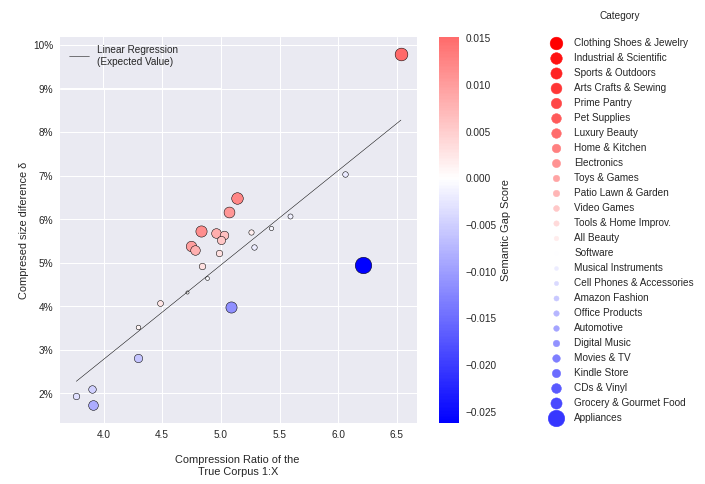}
\caption{Measures of the semantic gap between the language used in reviews and product descriptions. Greater intensity of the red and greater size of the mark indicate greater semantic gap. Greater intensity of the blue and greater size of the mark indicate a smaller semantic gap. \label{fig:scatter}}
\end{figure}

\section{Results}\label{sec:results}

Here we present the results of applying the method to measure the semantic gap (subsection \ref{sec:semantic_gap}) in each of the 28 domains of the Amazon dataset. Although the validity of the presented method is theoretically based on the distributional hypothesis for word semantics and on information theory, the experiments seek to provide additional experimental validation. For this we use the word embedding method (see subsection \ref{sec:drift_embeddings}), which is already known to be suitable for the task of measuring semantic distances between words, provided that two large and balanced corpora are available. As Table \ref{corpus_data} shows, the data set has large differences between the sizes of each corpus on each domain, as well as the size imbalance between reviews and descriptions. Consequently, the likely poor results obtained from these small and highly unbalanced corpus using word embeddings will be manually curated to select only quality word drifts for valitating our semantic gap method. The results of this experiment provide the information to test $H_1$. For the $H_2$ test, the semantic gap measures for each domain are compared to the average of the numerical ratings associated with each product review for each domain.

Figure \ref{fig:scatter} shows the relative differences obtained between the corpus \emph{True} and \emph{Rand} for each domain plotted against the compression ratio of \emph{True}. Since these differences are correlated with the compression ratio, the deviations from the trend line reflect the magnitude of the semantic gap between reviews and descriptions. Table \ref{tab:results} reports the magnitude of these deviations in the column ``Gap score'', which represents our function to measure the semantic gap, also used to order the domains in Table \ref{tab:results}.

To check the validity of the variable ``Gap score'', we use the ground truth described in subsection \ref{sec:groundtruth}. For each domain, we average the Jaccard scores (from eq. \ref{eq:score}) for the leading words according to $S()$ until all ten words in the ground truth are retrieved. These values are reported in Table \ref{tab:results} in the column labeled ``Avg.J@10''. We compare this with the ``Gap score'' and carefully check for other potential confounders, such as the size of the corpus of reviews or descriptions and their differences, to avoid a spurious correlation. Since ``Avg.J@10'' failed by far the \citet{d1973tests}\footnote{We used this implementation of the test: \url{https://docs.scipy.org/doc/scipy/reference/generated/scipy.stats.normaltest.html}} normality test ($p=4.5\times 10^{-8}$), we opted for the Spearman correlation as a non-parametric alternative obtaining $\rho=0.49$, $n =26$ ($p<0.01)$. This result proves the hypothesis $H_{1}$. Regarding $H_0^{aux}$, we do not find any supporting evidence because  all differences in the sizes of the compressed files $\delta$ for each domain were positive and the expert annotator found valid word drifts in all domains. This means that all studied domains exhibits some degree of semantical gap, leading us to reject $H_0^{aux}$ according to the available data.

To evaluate the relationship between the `` Gap score '' and customer satisfaction, we obtained the average of the product ratings on a 5-starts scale associated with the reviews having the attribute ``verified''  in the dataset for each domain. For this, we report in Table \ref{tab:results} the number of verified reviews along with the average rating given by customers along with their standard deviation. The independent variable ``Gap Score'' is continuous and normally distributed, since it passed the \citet{d1973tests} test  ($p=0.052>0.05$). Similarly, the dependent variable ``Avg. Rating'' is continuous, since it is the average of numerous discrete ratings in each domain, and it is also normally distributed as demonstrated by the same test ($p=0.257>0.05$). Thus, the Pearson correlation seems appropriate to measure the relationship between those variables, obtaining $r=0.32$ for $n =28$ ($p<0.05$). This result tests the hypothesis $H_{2}$. As the normality test of the ``Gap score'' variable is close to the critical value, we also considered the Spearman correlation, obtaining $\rho=0.28$, ($p<0.10)$. Although with less statistical significance, this result reconfirms the $H_{2}$ test.

\begin{table}[htb]

\caption{\label{corpus} Results of our semantic gap function along with the average rating (standard deviation in parentheses) and average Jaccard index for the words with the largest drift (until the 10th ground truth word was retrieved)\label{tab:results}}

{\small
\begin{tabular}{lrrrrr}
\hline \textbf{Category domain} & \textbf{Gap rank} & \textbf{Gap Score} & \textbf{Verif. Rev.} & \textbf{Avg. rating} & \textbf{Avg.J@10}\\ \hline
Clothing Shoes \& Jewelry&1&0.0151&30,286,706&4.19(1.23)&0.7306\\
Industrial \& Scientific&2&0.0124&1,636,564&4.32(1.22)&0.6529\\
Sports \& Outdoors&3&0.0115&11,913,059&4.26(1.23)&0.4781\\
Arts Crafts \& Sewing&4&0.0107&2,682,955&4.33(1.20)&3.2946\\
Prime Pantry&5&0.0099&382,686&4.38(1.15)&1.1008\\
Pet Supplies&6&0.0083&5,979,680&4.17(1.32)&1.0101\\
Luxury Beauty&7&0.0082&504,542&4.25(1.29)&2.1483\\
Home \& Kitchen&8&0.0064&20,015,942&4.22(1.29)&0.7359\\
Electronics&9&0.0058&18,597,092&4.11(1.36)&1.3168\\
Toys \& Games&10&0.0033&7,231,522&4.26(1.25)&0.6669\\
Patio Lawn \& Garden&11&0.0032&4,799,516&4.15(1.35)&0.9936\\
Video Games&12&0.0025&1,948,309&4.10(1.37)&0.2566\\
Tools \& Home Improvement&13&0.0018&8,299,068&4.25(1.27)&0.5611\\
All Beauty&14&0.0011&322,473&4.11(1.36)&1.5279\\
Software&15&0.0000&309,345&3.75(1.56)&2.0888\\
Musical Instruments&16&-0.0005&1,346,124&4.28(1.22)&0.6349\\
Cell Phones \& Accessories&17&-0.0008&9,209,864&3.94(1.46)&2.2113\\
Amazon Fashion&18&-0.0016&828,699&3.90(1.42)&2.5973\\
Office Products&19&-0.0020&5,0551,52&4.22(1.30)&1.0930\\
Automotive&20&-0.0021&7,561,414&4.27(1.27)&1.0790\\
Digital Music&21&-0.0033&1,217,667&4.69(0.79)&1.9304\\
Movies \& TV&22&-0.0047&6,731,296&4.31(1.17)&1.5674\\
Kindle Store&23&-0.0060&4,036,164&4.28(1.11)&1.9401\\
CDs \& Vinyl&24&-0.0085&2,578,257&4.60(0.88)&2.6809\\
Grocery \& Gourmet Food&25&-0.0114&4,437,360&4.35(1.21)&1.8710\\
Magazine Subscriptions&26&-0.0164&58,654&4.21(1.29)&n.a.\\
Gift Cards&27&-0.0233&138,237&4.71(0.88)&n.a.\\
Appliances&28&-0.0262&563,870&4.35(1.23)&6.8799\\
\hline
\end{tabular}
}
\end{table}

\section{Discussion}\label{sec:discussion}
 In the results obtained, we observe that domains whose items are related in some way to the language got a small gap, e.g., \emph{Magazine Subscriptions}, \emph{CDs \& Vinyl}, \emph{Kindle Store}, and \emph{Movies \& TV}. Not surprisingly, the \emph{Appliances} domain, an industry with a long tradition of user-friendly product descriptions, has the smallest semantic gap, implying well-aligned language between customers and suppliers.  In contrast, more technical domains such as \emph{Industrial \& Scientific}, \emph{Electronics}, and \emph{Sports \& Outdoors} had more significant semantic gaps. However, other technical domains such as \emph{Software} and \emph{Video Games} obtained an intermediate gap, showing a better alignment in the customers' language and that of the descriptions writers. Although the relationship between customer satisfaction and the quality of the descriptions has already been widely supported in the related literature \citep{ho1999antecedents,limayem2000makes,jiang2013measuring,lee2014consumers}, this result shows that the quality of the information is not the only factor, but rather that the use of a language semantically close to that of customers is also important.
 
 In general, we consider that the semantic gaps obtained for each domain are correct, which shows that the method proposed in subsection \ref{sec:semantic_gap} is a plausible approximation to the problem. These observations also provide evidence that the information-theoretical approach is sound and that a commonly used file compressor, such as \emph{bzip2}, is a suitable tool for the task.
 
 The observed correlation between the semantic gap and the average review ratings is essential because ratings can be associated with overall customer satisfaction \cite{engler2015understanding}. Thus, the average customer ratings reflect their level of satisfaction, whether it be for receiving better recommendations, exceeding expectations, convenient user interface, product quality, descriptions quality, or other factors. Our findings show that one of these factors may be the narrow semantic gap between the language of the customer community and the textual content produced by the provider. It is well known that customer satisfaction has a favorable impact on e-commerce trust \cite{kassim2010effect}. If this relationship can be extended to this linguistic factor, which has the potential to be controllable, it opens up a clear perspective for improving customers' shopping experience and, ultimately, business profits. This new relationship between the linguistic-semantic alignment of customer-provider communication and customer satisfaction, along with the considerable variation observed in this linguistic factor across different domains, provides tools for managers to make differential, revenue-weighted intervention decisions. Managerial decisions aimed at improving customer satisfaction are generally differentiated, for example, in the types of customers, in order to optimize the resources invested \citep{keiningham2005does}. The semantic gap measure provides the information necessary to support these decisions, either for different domains or for any other categorization that can be performed in the input corpus. The possible measures that can be applied to reduce the semantic gap could be to follow the recommendations of the "plain language" movement \citep{petelin2010considering}, mainly taking into account the intended audience of each document directed to the customers that the organization produces. From a theoretical point of view, the discovery of the direct relationship between the absence of semantic gap and customer satisfaction provides an important resource for developing behavior models, where the use of language by customers is an independent and observable variable, while on the supplier side it is controllable.
 
 Regarding the relationship between this semantic gap and the quality of customer-to-customer communication in social commerce, our results do not provide direct evidence of its existence. However, it is plausible to infer that customer satisfaction is also related to better communication due to less ambiguous language.
 
 Regarding the observed relationship between our semantic gap function and human judgments of the ground truth (``Avg.J@10'' in Table \ref{tab:results}), this is a quantitative indication that the information-theoretical method presented measures the semantic gap due to word drift and ambiguity. It can be concluded that the  method manages to effectively measure the semantic gap between customers and suppliers in e-commerce, but has the limitation that it does not identify particular words. However, the manually extracted ground truth is the necessary resource for developing and evaluating the methods to identify these words.

One of the advantages of the proposed method for measuring the semantic gap between customers and suppliers is that it is not based on particular features of the English language. Therefore, it can be applied in any language where the speech can be segmented into words. Regarding the required size of the corpus of reviews and descriptions, the Amazon data provided a wide range of sizes, but even the smallest corpus (see Table \ref{corpus_data}) can be considered large compared to those that can be collected from other e-commerce sites or from other smaller domains. Although we do not provide evidence of the reliability of the semantic gap measurements for small corpus, we consider that a corpus of reviews with more than 1 million words and a corpus of product descriptions with at least 100,000 words for each domain should be sufficient for a reliable measurement.

\section{Conclusions, Limitations, and Implications}\label{sec:conclusions}

We present a new method motivated by \textcolor{black}{distributional semantics and} information theory based on file compression to measure the lexical-semantic gap between product reviews and descriptions for various e-commerce domains. The proposed computational linguistics method makes very few assumptions, has no parameters, is language independent, and it do not require manual annotations or labor. It is based on Shannon's information theory and draws inspiration from other applications where this approach has been effective. In addition, by using word embeddings, we build a ground truth for the task of identifying words with semantic drift (i.e., different meanings) between product reviews and descriptions. This ground truth served to validate the soundness of the presented method. To obtain this gold standard, we applied a state-of-the-art natural language processing method (i.e. neural word embeddings) that produced a result that was selected manually by a professional linguist. This new resource provided additional validation, since the semantic gap measures for each domain ($n=26$) were highly correlated ($r=0.67$) with the average semantic drifts of the words in the semi-automatically obtained model. Furthermore, this gold standard constitutes a new resource for future research on this phenomenon. 

The most important observation of this study is that the semantic gap of various e-commerce domains has an inverse relationship with the quantitative ratings (on a 5-star scale) that customers give to the products purchased. This implies that the semantic differences in the language of the customers and suppliers have a relationship with customer satisfaction. Thus, the greater the semantic gap, the lower the customer satisfaction, represented by their ratings. This relationship was observed using the proposed method in a large corpus consisting of 6.9 billion words of user reviews and 1.5 billion words of product descriptions. This corpus came from Amazon and is distributed across 28 marketing domains. The observed correlation between the semantic gap with customer satisfaction represented by the average of the verified ratings was $r=0.32$ ($p\leq 0.05$). This data is probably the largest and most representative social e-commerce data publicly available.

\subsection{Theoretical Contribution}
Customer satisfaction in an e-commerce environment is a widely studied multifactorial phenomenon, but with few studies related to language. In social e-commerce, language plays a more prominent role than in classic e-commerce environments due to the written dialogue between customers. The differences in the meaning of words in the discourse of customers and suppliers, what we have called the \emph{semantic gap}, is a new factor that we show and that can be effectively measured even with the limitations of the scenario. The motivation for revealing this non-superficial linguistic feature is that semantics is probably the main factor associated with the effectiveness of human communication. We have found that the semantic gap is negatively related to customer satisfaction. This means that language not only has a functional-utilitarian role in the customer-provider relationship, but it also affects the quality of this relationship with its known implications for the business.

The semantic gap is a new variable that adds to those already known as explanatory of the phenomenon of customer satisfaction such as logistics quality, product quality, customer service, personalization, etc. This new variable can contribute to improving the predictive capacity of customer satisfaction models, since it belongs to a domain external to management, and that is generally not considered in current theory.

This contribution is of practical use in any social e-commerce environment where there is considerable availability of customer-written texts (e.g., reviews) and textual descriptions of products. In this scenario, the semantic gap provides a quantitative measure to identify subsets of customers with low satisfaction due to this linguistic difference.

\subsection{Limitations}

From the linguistic point of view, the method presented to measure the semantic gap has very few limitations to be applied in other types of texts or in other languages. This is because, in practice, it is only required to be able to segment texts at the word level, perform random manipulations at this level, and then use a standard file compressor. In agglutinative languages (e.g. Finnish) it may be necessary to perform word segmentation to control excessive vocabulary size. With regard to other types of documents to represent users (e.g. complaints and requests) and providers (e.g. manuals and regulations), it is reasonable to consider that $H_1$ may also be compatible with these types of data. Similarly, the semantic gap method could also be used with transcripts of dialogues in oral modality.

A limitation of this study is that its applicability is clear for large retail companies that have a considerable number of domains to compare. In e-commerce based on a single domain, the semantic gap measure loses its usefulness if it is not comparable with other domains, since it is relative to factors such as the size of the corpus and the compression ratio. However, relative semantic gap measures could be used in customer groupings other than by product domain, such as demographic or behavioral factors, among others.\\

Although it is plausible and appealing to generalize that a small semantic gap, which may imply good communication, could be positively related to customer satisfaction, our results show that this relationship, although statistically significant, is relatively weak. Consequently, this relationship should be considered with caution in other domains not very similar to those studied, such as finance, real estate, among others. Similarly, the possible causality from semantic gap to customer satisfaction seems reasonable, because of the apparent impossibility of an implication in the opposite direction. However, this consideration should be done with caution since our results outline possible confounding factors such as the degree to which language is involved with the product, as in the case of magazines and movies.

\subsection{Implications}
The main contributions and implications of this study can be summarized as:
\begin{enumerate}
\item For managers: For different marketing domains, there are varying degrees of differences in meaning between the words used by customers and those used in product descriptions (i.e. semantic gap). This difference is related to customer satisfaction: the smaller the semantic gap, the higher the customer satisfaction. This result was obtained without assumptions or information different from the linguistic data associated with each domain, so it is reasonable that it can be transferred beyond the 28 Amazon domains studied. To measure such a gap, it is only necessary to have a relatively large corpus of texts produced by customers (for example, product reviews) and another corpus with product descriptions. These two corpuses are compared by a simple preprocessing (accessible to any professional computer programmer) and with the use of a standard file compressor such as bzip2. Any action in the written discourse in the product descriptions aimed at reducing the semantic gap is expected to have a beneficial effect on customer satisfaction and ultimately on business profits.
\item For academicians: The relationship between the quality of information in product descriptions and customer satisfaction is well known. This study has specified the generality of the concept of "quality of information" to that of "quality of communication" between customers and suppliers when product descriptions are used as a communication channel. Here, the quality of communication refers to a low degree of ambiguity in the common terminology used by customers and suppliers.
\item For computational linguists: The method to measure the semantic gap between two corpus by compression is a new approach that has proven particularly useful in cases where there is a large imbalance between the size of the corpus to be compared. The method can be easily adapted to measure the semantic gap in subsets of vocabulary words, even to measure the semantic drift of individual words. In addition, this method obtained results similar to those that can be obtained when comparing distances using neural word embeddings.
\end{enumerate}

\subsection{Future Work}
Several research perspectives are opened from the methods and resources presented in this study:
\begin{enumerate}
\item It is interesting to investigate how the lexical-semantic gap affects other factors of e-commerce, such as customer service, trust, loyalty, and others.
\item It is also interesting to study whether the semantic gap also affects customer-customer communications in social commerce and its implications. 
\item The users' reviews have other features in addition to their sentiment polarity (i.e., rating), such as intensity and usefulness, that can serve to explore the semantic gap and thus expand the understanding of it and its relationships with other factors of marketing.
\item Research on the possible transfer of the methods presented to other analogous domains, such as those of the citizen-government or patient-healthcare system, becomes relevant.
\end{enumerate}


\bibliographystyle{model1-num-names}
\bibliography{main}

\begin{thebibliography}{61}
\expandafter\ifx\csname natexlab\endcsname\relax\def\natexlab#1{#1}\fi
\providecommand{\url}[1]{\texttt{#1}}
\providecommand{\href}[2]{#2}
\providecommand{\path}[1]{#1}
\providecommand{\DOIprefix}{doi:}
\providecommand{\ArXivprefix}{arXiv:}
\providecommand{\URLprefix}{URL: }
\providecommand{\Pubmedprefix}{pmid:}
\providecommand{\doi}[1]{\href{http://dx.doi.org/#1}{\path{#1}}}
\providecommand{\Pubmed}[1]{\href{pmid:#1}{\path{#1}}}
\providecommand{\bibinfo}[2]{#2}
\ifx\xfnm\relax \def\xfnm[#1]{\unskip,\space#1}\fi
\bibitem[{Kim and Lee(2002)}]{kim2002critical}
\bibinfo{author}{J.~Kim}, \bibinfo{author}{J.~Lee},
\newblock \bibinfo{title}{Critical design factors for successful e-commerce
  systems},
\newblock \bibinfo{journal}{Behaviour \& Information Technology}
  \bibinfo{volume}{21} (\bibinfo{year}{2002}) \bibinfo{pages}{185--199}.
\bibitem[{Hajli(2019)}]{hajli2019impact}
\bibinfo{author}{N.~Hajli},
\newblock \bibinfo{title}{The impact of positive valence and negative valence
  on social commerce purchase intention},
\newblock \bibinfo{journal}{Information Technology \& People}
  (\bibinfo{year}{2019}).
\bibitem[{Adler(2012)}]{adler2012plain}
\bibinfo{author}{M.~Adler},
\newblock \bibinfo{title}{The plain language movement},
\newblock in: \bibinfo{booktitle}{The Oxford handbook of language and law},
  \bibinfo{year}{2012}.
\bibitem[{Vengadasamy et~al.(2004)Vengadasamy, Jaludin, and
  Hamat}]{vengadasamy2004characteristics}
\bibinfo{author}{R.~Vengadasamy}, \bibinfo{author}{A.~Jaludin},
  \bibinfo{author}{A.~Hamat},
\newblock \bibinfo{title}{Characteristics of written text in e-commerce
  websites},
\newblock \bibinfo{journal}{Internet Journal of e-Language Learning \&
  Teaching} \bibinfo{volume}{1} (\bibinfo{year}{2004}) \bibinfo{pages}{15--32}.
\bibitem[{Bingi et~al.(2006)Bingi, Mir, and Khamalah}]{bingi2006challenges}
\bibinfo{author}{P.~Bingi}, \bibinfo{author}{A.~Mir},
  \bibinfo{author}{J.~Khamalah},
\newblock \bibinfo{title}{The challenges facing global e-commerce},
\newblock \bibinfo{journal}{Information Systems Management}
  (\bibinfo{year}{2006}).
\bibitem[{Barrera et~al.(2016)Barrera, Popescu, Toral, Gaspari, and
  Choukri}]{barrera2016enhancing}
\bibinfo{author}{M.~F. Barrera}, \bibinfo{author}{V.~Popescu},
  \bibinfo{author}{A.~Toral}, \bibinfo{author}{F.~Gaspari},
  \bibinfo{author}{K.~Choukri},
\newblock \bibinfo{title}{Enhancing cross-border eu e-commerce through machine
  translation: Needed language resources, challenges and opportunities},
\newblock in: \bibinfo{booktitle}{Proceedings of the Tenth International
  Conference on Language Resources and Evaluation (LREC'16)},
  \bibinfo{year}{2016}, pp. \bibinfo{pages}{4550--4556}.
\bibitem[{Rust and Zahorik(1993)}]{rust1993customer}
\bibinfo{author}{R.~T. Rust}, \bibinfo{author}{A.~J. Zahorik},
\newblock \bibinfo{title}{Customer satisfaction, customer retention, and market
  share},
\newblock \bibinfo{journal}{Journal of retailing} \bibinfo{volume}{69}
  (\bibinfo{year}{1993}) \bibinfo{pages}{193--215}.
\bibitem[{Eid(2011)}]{eid2011determinants}
\bibinfo{author}{M.~I. Eid},
\newblock \bibinfo{title}{Determinants of e-commerce customer satisfaction,
  trust, and loyalty in saudi arabia},
\newblock \bibinfo{journal}{Journal of electronic commerce research}
  \bibinfo{volume}{12} (\bibinfo{year}{2011}) \bibinfo{pages}{78}.
\bibitem[{Tripathi(2014)}]{tripathi2014customer}
\bibinfo{author}{M.~Tripathi},
\newblock \bibinfo{title}{Customer satisfaction and engagement-customer
  retention strategies for brand manager.},
\newblock \bibinfo{journal}{Vilakshan: The XIMB Journal of Management}
  \bibinfo{volume}{11} (\bibinfo{year}{2014}).
\bibitem[{Rese(2003)}]{rese2003relationship}
\bibinfo{author}{M.~Rese},
\newblock \bibinfo{title}{Relationship marketing and customer satisfaction: an
  information economics perspective},
\newblock \bibinfo{journal}{Marketing theory} \bibinfo{volume}{3}
  (\bibinfo{year}{2003}) \bibinfo{pages}{97--117}.
\bibitem[{Kassim and Abdullah(2010)}]{kassim2010effect}
\bibinfo{author}{N.~Kassim}, \bibinfo{author}{N.~A. Abdullah},
\newblock \bibinfo{title}{The effect of perceived service quality dimensions on
  customer satisfaction, trust, and loyalty in e-commerce settings},
\newblock \bibinfo{journal}{Asia pacific journal of marketing and logistics}
  (\bibinfo{year}{2010}).
\bibitem[{Biswas et~al.(2019)Biswas, Nusari, Ghosh
  et~al.}]{biswas2019influence}
\bibinfo{author}{K.~M. Biswas}, \bibinfo{author}{M.~Nusari},
  \bibinfo{author}{A.~Ghosh}, et~al.,
\newblock \bibinfo{title}{The influence of website service quality on customer
  satisfaction towards online shopping: The mediating role of confirmation of
  expectation},
\newblock \bibinfo{journal}{International Journal of Management Science and
  Business Administration} \bibinfo{volume}{5} (\bibinfo{year}{2019})
  \bibinfo{pages}{7--14}.
\bibitem[{Hua and Jing(2015)}]{hua2015empirical}
\bibinfo{author}{W.~Hua}, \bibinfo{author}{Z.~Jing},
\newblock \bibinfo{title}{An empirical study on e-commerce logistics service
  quality and customer satisfaction},
\newblock \bibinfo{journal}{WHICEB Proceeding}  (\bibinfo{year}{2015})
  \bibinfo{pages}{269--275}.
\bibitem[{Skordoulis et~al.(2018)Skordoulis, Kaskouta, Chalikias, and
  Drosos}]{skordoulis2018commerce}
\bibinfo{author}{M.~Skordoulis}, \bibinfo{author}{I.~Kaskouta},
  \bibinfo{author}{M.~Chalikias}, \bibinfo{author}{D.~Drosos},
\newblock \bibinfo{title}{E-commerce and e-customer satisfaction during the
  economic crisis},
\newblock \bibinfo{journal}{Journal for International Business and
  Entrepreneurship Development} \bibinfo{volume}{11} (\bibinfo{year}{2018})
  \bibinfo{pages}{15--29}.
\bibitem[{Gong-min(2010)}]{gong2010research}
\bibinfo{author}{Z.~Gong-min},
\newblock \bibinfo{title}{Research on customer loyalty of b2c e-commerce},
\newblock \bibinfo{journal}{China-USA Business Review} \bibinfo{volume}{9}
  (\bibinfo{year}{2010}) \bibinfo{pages}{46}.
\bibitem[{Mu et~al.(2021)Mu, Zheng, Zhang, and Zhang}]{mu2021research}
\bibinfo{author}{R.~Mu}, \bibinfo{author}{Y.~Zheng},
  \bibinfo{author}{K.~Zhang}, \bibinfo{author}{Y.~Zhang},
\newblock \bibinfo{title}{Research on customer satisfaction based on
  multidimensional analysis},
\newblock \bibinfo{journal}{International Journal of Computational Intelligence
  Systems} \bibinfo{volume}{14} (\bibinfo{year}{2021})
  \bibinfo{pages}{605--616}.
\bibitem[{Pryzant et~al.(2017)Pryzant, Chung, and
  Jurafsky}]{pryzant2017predicting}
\bibinfo{author}{R.~Pryzant}, \bibinfo{author}{Y.~Chung},
  \bibinfo{author}{D.~Jurafsky},
\newblock \bibinfo{title}{Predicting sales from the language of product
  descriptions},
\newblock in: \bibinfo{booktitle}{eCOM@ SIGIR}, \bibinfo{year}{2017}.
\bibitem[{Elsholz et~al.(2019)Elsholz, Chamberlain, and
  Kruschwitz}]{elsholz2019exploring}
\bibinfo{author}{E.~Elsholz}, \bibinfo{author}{J.~Chamberlain},
  \bibinfo{author}{U.~Kruschwitz},
\newblock \bibinfo{title}{Exploring language style in chatbots to increase
  perceived product value and user engagement},
\newblock in: \bibinfo{booktitle}{Proceedings of the 2019 Conference on Human
  Information Interaction and Retrieval}, \bibinfo{year}{2019}, pp.
  \bibinfo{pages}{301--305}.
\bibitem[{Beinborn and Choenni(2020)}]{beinborn2020semantic}
\bibinfo{author}{L.~Beinborn}, \bibinfo{author}{R.~Choenni},
\newblock \bibinfo{title}{Semantic drift in multilingual representations},
\newblock \bibinfo{journal}{Computational Linguistics} \bibinfo{volume}{46}
  (\bibinfo{year}{2020}) \bibinfo{pages}{571--603}.
\bibitem[{Piantadosi et~al.(2012)Piantadosi, Tily, and
  Gibson}]{piantadosi2012communicative}
\bibinfo{author}{S.~T. Piantadosi}, \bibinfo{author}{H.~Tily},
  \bibinfo{author}{E.~Gibson},
\newblock \bibinfo{title}{The communicative function of ambiguity in language},
\newblock \bibinfo{journal}{Cognition} \bibinfo{volume}{122}
  (\bibinfo{year}{2012}) \bibinfo{pages}{280--291}.
\bibitem[{Firth(1957)}]{firth1957synopsis}
\bibinfo{author}{J.~R. Firth},
\newblock \bibinfo{title}{A synopsis of linguistic theory, 1930-1955},
\newblock \bibinfo{journal}{Studies in linguistic analysis}
  (\bibinfo{year}{1957}).
\bibitem[{Roberts(2016)}]{roberts2016assessing}
\bibinfo{author}{K.~Roberts},
\newblock \bibinfo{title}{Assessing the corpus size vs. similarity trade-off
  for word embeddings in clinical nlp},
\newblock in: \bibinfo{booktitle}{Proceedings of the Clinical Natural Language
  Processing Workshop (ClinicalNLP)}, \bibinfo{year}{2016}, pp.
  \bibinfo{pages}{54--63}.
\bibitem[{Salton et~al.(1975)Salton, Wong, and Yang}]{salton1975vector}
\bibinfo{author}{G.~Salton}, \bibinfo{author}{A.~Wong}, \bibinfo{author}{C.-S.
  Yang},
\newblock \bibinfo{title}{A vector space model for automatic indexing},
\newblock \bibinfo{journal}{Communications of the ACM} \bibinfo{volume}{18}
  (\bibinfo{year}{1975}) \bibinfo{pages}{613--620}.
\bibitem[{Sahlgren(2006)}]{sahlgren2006word}
\bibinfo{author}{M.~Sahlgren}, \bibinfo{title}{The Word-Space Model: Using
  distributional analysis to represent syntagmatic and paradigmatic relations
  between words in high-dimensional vector spaces}, Ph.D. thesis, Stockholm
  University, \bibinfo{year}{2006}.
\bibitem[{Deerwester et~al.(1990)Deerwester, Dumais, Furnas, Landauer, and
  Harshman}]{deerwester1990indexing}
\bibinfo{author}{S.~Deerwester}, \bibinfo{author}{S.~T. Dumais},
  \bibinfo{author}{G.~W. Furnas}, \bibinfo{author}{T.~K. Landauer},
  \bibinfo{author}{R.~Harshman},
\newblock \bibinfo{title}{Indexing by latent semantic analysis},
\newblock \bibinfo{journal}{Journal of the American society for information
  science} \bibinfo{volume}{41} (\bibinfo{year}{1990})
  \bibinfo{pages}{391--407}.
\bibitem[{Turian et~al.(2010)Turian, Ratinov, and Bengio}]{turian2010word}
\bibinfo{author}{J.~Turian}, \bibinfo{author}{L.~Ratinov},
  \bibinfo{author}{Y.~Bengio},
\newblock \bibinfo{title}{Word representations: a simple and general method for
  semi-supervised learning},
\newblock in: \bibinfo{booktitle}{Proceedings of the 48th annual meeting of the
  association for computational linguistics}, \bibinfo{year}{2010}, pp.
  \bibinfo{pages}{384--394}.
\bibitem[{Mikolov et~al.(2013)Mikolov, Sutskever, Chen, Corrado, and
  Dean}]{mikolov2013distributed}
\bibinfo{author}{T.~Mikolov}, \bibinfo{author}{I.~Sutskever},
  \bibinfo{author}{K.~Chen}, \bibinfo{author}{G.~Corrado},
  \bibinfo{author}{J.~Dean},
\newblock \bibinfo{title}{Distributed representations of words and phrases and
  their compositionality},
\newblock in: \bibinfo{booktitle}{Proceedings NIPS}, \bibinfo{year}{2013}, p.
  \bibinfo{pages}{pages 3111–3119}.
\bibitem[{Pennington et~al.(2014)Pennington, Socher, and
  Manning}]{pennington2014glove}
\bibinfo{author}{J.~Pennington}, \bibinfo{author}{R.~Socher},
  \bibinfo{author}{C.~D. Manning},
\newblock \bibinfo{title}{Glove: Global vectors for word representation},
\newblock in: \bibinfo{booktitle}{Proceedings of the 2014 conference on
  empirical methods in natural language processing (EMNLP)},
  \bibinfo{year}{2014}, pp. \bibinfo{pages}{1532--1543}.
\bibitem[{Zheng et~al.(2013)Zheng, Zhu, and Lin}]{zheng2013capturing}
\bibinfo{author}{X.~Zheng}, \bibinfo{author}{S.~Zhu}, \bibinfo{author}{Z.~Lin},
\newblock \bibinfo{title}{Capturing the essence of word-of-mouth for social
  commerce: Assessing the quality of online e-commerce reviews by a
  semi-supervised approach},
\newblock \bibinfo{journal}{Decision Support Systems} \bibinfo{volume}{56}
  (\bibinfo{year}{2013}) \bibinfo{pages}{211--222}.
\bibitem[{Shannon(1948)}]{shannon1948mathematical}
\bibinfo{author}{C.~E. Shannon},
\newblock \bibinfo{title}{A mathematical theory of communication},
\newblock \bibinfo{journal}{The Bell system technical journal}
  \bibinfo{volume}{27} (\bibinfo{year}{1948}) \bibinfo{pages}{379--423}.
\bibitem[{Francis et~al.(2002)Francis, White et~al.}]{francis2002pirqual}
\bibinfo{author}{J.~E. Francis}, \bibinfo{author}{L.~White}, et~al.,
\newblock \bibinfo{title}{Pirqual: a scale for measuring customer expectations
  and perceptions of quality in internet retailing},
\newblock \bibinfo{journal}{K. Evans \& L. Scheer (Eds.)}
  (\bibinfo{year}{2002}) \bibinfo{pages}{263--270}.
\bibitem[{Ludin and Cheng(2014)}]{ludin2014factors}
\bibinfo{author}{I.~H. B.~H. Ludin}, \bibinfo{author}{B.~L. Cheng},
\newblock \bibinfo{title}{Factors influencing customer satisfaction and
  e-loyalty: Online shopping environment among the young adults},
\newblock \bibinfo{journal}{Management Dynamics in the Knowledge Economy}
  \bibinfo{volume}{2} (\bibinfo{year}{2014}) \bibinfo{pages}{462}.
\bibitem[{Patrada and Andajani(2021)}]{patrada2021effect}
\bibinfo{author}{R.~Patrada}, \bibinfo{author}{E.~Andajani},
\newblock \bibinfo{title}{Effect and consequence e-customer satisfaction for
  e-commerce users},
\newblock \bibinfo{journal}{IPTEK Journal of Proceedings Series}
  (\bibinfo{year}{2021}) \bibinfo{pages}{219--227}.
\bibitem[{Mou et~al.(2019)Mou, Zhu, and Benyoucef}]{mou2019impact}
\bibinfo{author}{J.~Mou}, \bibinfo{author}{W.~Zhu},
  \bibinfo{author}{M.~Benyoucef},
\newblock \bibinfo{title}{Impact of product description and involvement on
  purchase intention in cross-border e-commerce},
\newblock \bibinfo{journal}{Industrial Management \& Data Systems}
  (\bibinfo{year}{2019}).
\bibitem[{Wang et~al.(2018)Wang, Li, Tian, Wang, and
  Cheng}]{wang2018extracting}
\bibinfo{author}{W.~M. Wang}, \bibinfo{author}{Z.~Li},
  \bibinfo{author}{Z.~Tian}, \bibinfo{author}{J.~Wang},
  \bibinfo{author}{M.~Cheng},
\newblock \bibinfo{title}{Extracting and summarizing affective features and
  responses from online product descriptions and reviews: A kansei text mining
  approach},
\newblock \bibinfo{journal}{Engineering Applications of Artificial
  Intelligence} \bibinfo{volume}{73} (\bibinfo{year}{2018})
  \bibinfo{pages}{149--162}.
\bibitem[{Novgorodov et~al.(2019)Novgorodov, Guy, Elad, and
  Radinsky}]{novgorodov2019generating}
\bibinfo{author}{S.~Novgorodov}, \bibinfo{author}{I.~Guy},
  \bibinfo{author}{G.~Elad}, \bibinfo{author}{K.~Radinsky},
\newblock \bibinfo{title}{Generating product descriptions from user reviews},
\newblock in: \bibinfo{booktitle}{The World Wide Web Conference},
  \bibinfo{year}{2019}, pp. \bibinfo{pages}{1354--1364}.
\bibitem[{Zhang et~al.(2019)Zhang, Zhang, Huo, and Ren}]{zhang2019automatic}
\bibinfo{author}{T.~Zhang}, \bibinfo{author}{J.~Zhang},
  \bibinfo{author}{C.~Huo}, \bibinfo{author}{W.~Ren},
\newblock \bibinfo{title}{Automatic generation of pattern-controlled product
  description in e-commerce},
\newblock in: \bibinfo{booktitle}{The World Wide Web Conference},
  \bibinfo{year}{2019}, pp. \bibinfo{pages}{2355--2365}.
\bibitem[{Chen et~al.(2019)Chen, Lin, Zhang, Yang, Zhou, and
  Tang}]{chen2019towards}
\bibinfo{author}{Q.~Chen}, \bibinfo{author}{J.~Lin},
  \bibinfo{author}{Y.~Zhang}, \bibinfo{author}{H.~Yang},
  \bibinfo{author}{J.~Zhou}, \bibinfo{author}{J.~Tang},
\newblock \bibinfo{title}{Towards knowledge-based personalized product
  description generation in e-commerce},
\newblock in: \bibinfo{booktitle}{Proceedings of the 25th ACM SIGKDD
  International Conference on Knowledge Discovery \& Data Mining},
  \bibinfo{year}{2019}, pp. \bibinfo{pages}{3040--3050}.
\bibitem[{Elad et~al.(2019)Elad, Radinsky, and Kimelfeld}]{elad2019generating}
\bibinfo{author}{G.~Elad}, \bibinfo{author}{K.~Radinsky},
  \bibinfo{author}{B.~Kimelfeld}, \bibinfo{title}{Generating Personalized
  Product Descriptions from User Reviews}, Ph.D. thesis, Computer Science
  Department, Technion, \bibinfo{year}{2019}.
\bibitem[{Cilibrasi and Vitanyi(2003)}]{cilibrasi2003clustering}
\bibinfo{author}{R.~Cilibrasi}, \bibinfo{author}{P.~Vitanyi},
\newblock \bibinfo{title}{Clustering by compression},
\newblock \bibinfo{journal}{arXiv preprint cs/0312044}  (\bibinfo{year}{2003}).
\bibitem[{Montemurro and Zanette(2011)}]{montemurro2011universal}
\bibinfo{author}{M.~A. Montemurro}, \bibinfo{author}{D.~H. Zanette},
\newblock \bibinfo{title}{Universal entropy of word ordering across linguistic
  families},
\newblock \bibinfo{journal}{PLoS One} \bibinfo{volume}{6}
  (\bibinfo{year}{2011}) \bibinfo{pages}{e19875}.
\bibitem[{Ruder et~al.(2019)Ruder, Vuli{\'c}, and S{\o}gaard}]{ruder2019survey}
\bibinfo{author}{S.~Ruder}, \bibinfo{author}{I.~Vuli{\'c}},
  \bibinfo{author}{A.~S{\o}gaard},
\newblock \bibinfo{title}{A survey of cross-lingual word embedding models},
\newblock \bibinfo{journal}{Journal of Artificial Intelligence Research}
  \bibinfo{volume}{65} (\bibinfo{year}{2019}) \bibinfo{pages}{569--631}.
\bibitem[{Artetxe et~al.(2018)Artetxe, Labaka, and Agirre}]{artetxe2018robust}
\bibinfo{author}{M.~Artetxe}, \bibinfo{author}{G.~Labaka},
  \bibinfo{author}{E.~Agirre},
\newblock \bibinfo{title}{A robust self-learning method for fully unsupervised
  cross-lingual mappings of word embeddings},
\newblock \bibinfo{journal}{arXiv preprint arXiv:1805.06297}
  (\bibinfo{year}{2018}).
\bibitem[{Han et~al.(2018)Han, Gill, Spirling, and Cho}]{han2018conditional}
\bibinfo{author}{R.~Han}, \bibinfo{author}{M.~Gill},
  \bibinfo{author}{A.~Spirling}, \bibinfo{author}{K.~Cho},
\newblock \bibinfo{title}{Conditional word embedding and hypothesis testing via
  bayes-by-backprop},
\newblock in: \bibinfo{booktitle}{Proceedings of the 2018 Conference on
  Empirical Methods in Natural Language Processing}, \bibinfo{year}{2018}, pp.
  \bibinfo{pages}{4890--4895}.
\bibitem[{Kulkarni et~al.(2015)Kulkarni, Al-Rfou, Perozzi, and
  Skiena}]{kulkarni2015statistically}
\bibinfo{author}{V.~Kulkarni}, \bibinfo{author}{R.~Al-Rfou},
  \bibinfo{author}{B.~Perozzi}, \bibinfo{author}{S.~Skiena},
\newblock \bibinfo{title}{Statistically significant detection of linguistic
  change},
\newblock in: \bibinfo{booktitle}{Proceedings of the 24th International
  Conference on World Wide Web}, \bibinfo{year}{2015}, pp.
  \bibinfo{pages}{625--635}.
\bibitem[{Gouws and S{\o}gaard(2015)}]{gouws2015simple}
\bibinfo{author}{S.~Gouws}, \bibinfo{author}{A.~S{\o}gaard},
\newblock \bibinfo{title}{Simple task-specific bilingual word embeddings},
\newblock in: \bibinfo{booktitle}{Proceedings of the 2015 Conference of the
  North American Chapter of the Association for Computational Linguistics:
  Human Language Technologies}, \bibinfo{year}{2015}, pp.
  \bibinfo{pages}{1386--1390}.
\bibitem[{Vulic and Moens(2015)}]{vulic2015bilingual}
\bibinfo{author}{I.~Vulic}, \bibinfo{author}{M.-F. Moens},
\newblock \bibinfo{title}{Bilingual word embeddings from non-parallel
  document-aligned data applied to bilingual lexicon induction},
\newblock in: \bibinfo{booktitle}{Proceedings of the 53rd Annual Meeting of the
  Association for Computational Linguistics (ACL 2015)},
  volume~\bibinfo{volume}{2}, \bibinfo{organization}{ACL; East Stroudsburg,
  PA}, \bibinfo{year}{2015}, pp. \bibinfo{pages}{719--725}.
\bibitem[{Frermann and Lapata(2016)}]{frermann2016bayesian}
\bibinfo{author}{L.~Frermann}, \bibinfo{author}{M.~Lapata},
\newblock \bibinfo{title}{A bayesian model of diachronic meaning change},
\newblock \bibinfo{journal}{Transactions of the Association for Computational
  Linguistics} \bibinfo{volume}{4} (\bibinfo{year}{2016})
  \bibinfo{pages}{31--45}.
\bibitem[{Basile et~al.(2020)Basile, Caputo, Caselli, Cassotti, and
  Varvara}]{basile2020diacr}
\bibinfo{author}{P.~Basile}, \bibinfo{author}{A.~Caputo},
  \bibinfo{author}{T.~Caselli}, \bibinfo{author}{P.~Cassotti},
  \bibinfo{author}{R.~Varvara},
\newblock \bibinfo{title}{Diacr-ita@ evalita2020: Overview of the evalita2020
  diachronic lexical semantics (diacr-ita) task},
\newblock in: \bibinfo{booktitle}{Proceedings of the 7th evaluation campaign of
  Natural Language Processing and Speech tools for Italian (EVALITA 2020),
  Online. CEUR. org}, \bibinfo{year}{2020}.
\bibitem[{Angel et~al.(2020)Angel, Rodriguez-Diaz, Gelbukh, and
  Jimenez}]{angel2020nlp}
\bibinfo{author}{J.~Angel}, \bibinfo{author}{C.~A. Rodriguez-Diaz},
  \bibinfo{author}{A.~Gelbukh}, \bibinfo{author}{S.~Jimenez},
\newblock \bibinfo{title}{Nlp-cic@ diacr-ita: Pos and neighbor based
  distributional models for lexical semantic change in diachronic italian
  corpora},
\newblock \bibinfo{journal}{arXiv preprint arXiv:2011.03755}
  (\bibinfo{year}{2020}).
\bibitem[{Ni et~al.(2019)Ni, Li, and McAuley}]{ni2019justifying}
\bibinfo{author}{J.~Ni}, \bibinfo{author}{J.~Li}, \bibinfo{author}{J.~McAuley},
\newblock \bibinfo{title}{Justifying recommendations using distantly-labeled
  reviews and fine-grained aspects},
\newblock in: \bibinfo{booktitle}{Proceedings of the 2019 Conference on
  Empirical Methods in Natural Language Processing and the 9th International
  Joint Conference on Natural Language Processing (EMNLP-IJCNLP)},
  \bibinfo{year}{2019}, pp. \bibinfo{pages}{188--197}.
\bibitem[{Miller(1995)}]{miller1995wordnet}
\bibinfo{author}{G.~A. Miller},
\newblock \bibinfo{title}{Wordnet: a lexical database for english},
\newblock \bibinfo{journal}{Communications of the ACM} \bibinfo{volume}{38}
  (\bibinfo{year}{1995}) \bibinfo{pages}{39--41}.
\bibitem[{{\v R}eh{\r u}{\v r}ek and Sojka(2010)}]{rehurek_lrec}
\bibinfo{author}{R.~{\v R}eh{\r u}{\v r}ek}, \bibinfo{author}{P.~Sojka},
\newblock \bibinfo{title}{{Software Framework for Topic Modelling with Large
  Corpora}},
\newblock in: \bibinfo{booktitle}{{Proceedings of the LREC 2010 Workshop on New
  Challenges for NLP Frameworks}}, \bibinfo{publisher}{ELRA},
  \bibinfo{address}{Valletta, Malta}, \bibinfo{year}{2010}, pp.
  \bibinfo{pages}{45--50}.
\bibitem[{D'Agostino and Pearson(1973)}]{d1973tests}
\bibinfo{author}{R.~D'Agostino}, \bibinfo{author}{E.~S. Pearson},
\newblock \bibinfo{title}{Tests for departure from normality. empirical results
  for the distributions of $b_2$ and $\sqrt{b_1}$},
\newblock \bibinfo{journal}{Biometrika} \bibinfo{volume}{60}
  (\bibinfo{year}{1973}) \bibinfo{pages}{613--622}.
\bibitem[{Ho and Wu(1999)}]{ho1999antecedents}
\bibinfo{author}{C.-F. Ho}, \bibinfo{author}{W.-H. Wu},
\newblock \bibinfo{title}{Antecedents of customer satisfaction on the internet:
  an empirical study of online shopping},
\newblock in: \bibinfo{booktitle}{Proceedings of the 32nd Annual Hawaii
  International Conference on Systems Sciences. 1999. HICSS-32. Abstracts and
  CD-ROM of Full Papers}, \bibinfo{organization}{IEEE}, \bibinfo{year}{1999},
  pp. \bibinfo{pages}{9--pp}.
\bibitem[{Limayem et~al.(2000)Limayem, Khalifa, and Frini}]{limayem2000makes}
\bibinfo{author}{M.~Limayem}, \bibinfo{author}{M.~Khalifa},
  \bibinfo{author}{A.~Frini},
\newblock \bibinfo{title}{What makes consumers buy from internet? a
  longitudinal study of online shopping},
\newblock \bibinfo{journal}{IEEE Transactions on systems, man, and
  Cybernetics-Part A: Systems and Humans} \bibinfo{volume}{30}
  (\bibinfo{year}{2000}) \bibinfo{pages}{421--432}.
\bibitem[{Jiang et~al.(2013)Jiang, Yang, and Jun}]{jiang2013measuring}
\bibinfo{author}{L.~A. Jiang}, \bibinfo{author}{Z.~Yang},
  \bibinfo{author}{M.~Jun},
\newblock \bibinfo{title}{Measuring consumer perceptions of online shopping
  convenience},
\newblock \bibinfo{journal}{Journal of Service management}
  (\bibinfo{year}{2013}).
\bibitem[{Lee and Shin(2014)}]{lee2014consumers}
\bibinfo{author}{E.-J. Lee}, \bibinfo{author}{S.~Y. Shin},
\newblock \bibinfo{title}{When do consumers buy online product reviews? effects
  of review quality, product type, and reviewer’s photo},
\newblock \bibinfo{journal}{Computers in human behavior} \bibinfo{volume}{31}
  (\bibinfo{year}{2014}) \bibinfo{pages}{356--366}.
\bibitem[{Engler et~al.(2015)Engler, Winter, and
  Schulz}]{engler2015understanding}
\bibinfo{author}{T.~H. Engler}, \bibinfo{author}{P.~Winter},
  \bibinfo{author}{M.~Schulz},
\newblock \bibinfo{title}{Understanding online product ratings: A customer
  satisfaction model},
\newblock \bibinfo{journal}{Journal of Retailing and Consumer Services}
  \bibinfo{volume}{27} (\bibinfo{year}{2015}) \bibinfo{pages}{113--120}.
\bibitem[{Keiningham et~al.(2005)Keiningham, Perkins-Munn, Aksoy, and
  Estrin}]{keiningham2005does}
\bibinfo{author}{T.~L. Keiningham}, \bibinfo{author}{T.~Perkins-Munn},
  \bibinfo{author}{L.~Aksoy}, \bibinfo{author}{D.~Estrin},
\newblock \bibinfo{title}{Does customer satisfaction lead to profitability? the
  mediating role of share-of-wallet},
\newblock \bibinfo{journal}{Managing Service Quality: An International Journal}
   (\bibinfo{year}{2005}).
\bibitem[{Petelin(2010)}]{petelin2010considering}
\bibinfo{author}{R.~Petelin},
\newblock \bibinfo{title}{Considering plain language: issues and initiatives},
\newblock \bibinfo{journal}{Corporate Communications: An International Journal}
  \bibinfo{volume}{15} (\bibinfo{year}{2010}) \bibinfo{pages}{205--216}.

\end{thebibliography}
\end{document}